\documentclass[journal]{IEEEtran}
\IEEEoverridecommandlockouts
\usepackage{bbding}
\usepackage{cite}
\usepackage{amsmath,amssymb,amsfonts}
\usepackage{amsthm}
\usepackage{algorithmic}
\usepackage{algorithm}
\usepackage{graphicx}
\usepackage{textcomp}
\usepackage{xcolor}
\usepackage{booktabs}
\usepackage{multirow}
\usepackage{makecell}
\usepackage{url}

\floatname{algorithm}{Algorithm}

\def\BibTeX{{\rm B\kern-.05em{\sc i\kern-.025em b}\kern-.08em
    T\kern-.1667em\lower.7ex\hbox{E}\kern-.125emX}}

\usepackage{tikz}
\usepackage{hyperref}

\definecolor{lime}{HTML}{A6CE39}
\DeclareRobustCommand{\orcidicon}{
	\begin{tikzpicture}
		\draw[lime, fill=lime] (0,0)
		circle[radius=0.16]
		node[white]{{\fontfamily{qag}\selectfont \tiny \.{I}D}}; 
	\end{tikzpicture}
	\hspace{-2mm}
}
\foreach \x in {A, ..., Z}{%
	\expandafter\xdef\csname orcid\x\endcsname{\noexpand\href{https://orcid.org/\csname orcidauthor\x\endcsname}{\noexpand\orcidicon}}
}

\begin{document}

\title{Counterfactual Public Opinion Projection for Intervention Decision-Making}
\author{Ruixing~Ren, Junhui~Zhao,~\IEEEmembership{Senior~Member,~IEEE}, Qiuping~Li, Xiaoke~Sun 
	\thanks{This work was supported in part by National Natural Science Foundation of China under Grant U25B2007, in part by the the Fundamental Research Funds for the central Universities under Grant 2025JBZX060, in part by National Engineering Research Center of System Technology for High-Speed	Railway and Urban Rail Transit under Grant 2024YJ255. (Corresponding author: Junhui Zhao.)}%
	
	\thanks{Ruixing Ren, Junhui Zhao are with the School of Electronic and Information Engineering, Beijing Jiaotong University, Beijing 100044, China. (e-mail: renruixing0604@163.com; junhuizhao@hotmail.com)

		Qiuping Li and Xiaoke Sun  are with the National Computer Network Emergency Response Technical Team/Coordination Center of China (CNCERT/CC), Beijing 100029, China (e-mail: qiupingli\_bj@163.com; xiaokesun@cert.org.cn).
		
	}
	
}

\maketitle

\begin{abstract}
Effective public opinion intervention requires forecasts of how the same event would evolve under alternative actions. Existing methods mainly capture observational associations and inadequately address treatment selection bias. The paper proposes a counterfactual public opinion projection framework for intervention decision making. Each intervention is represented by its timing, type, content and intensity. Causal triggering structures are inferred from propagation event sequences. Counterfactual trajectories of attention, sentiment and polarization are generated from adversarially balanced representations. Candidate interventions are then compared through repeated rollouts. Semisynthetic experiments yielded an overall F1 comparable to a direct structure learning baseline and a recall of 0.698 for causal edges linking semantically dissimilar content. Representation diagnostics showed that the gradient reversal layer reduced treatment information leakage from 0.163 to 0.140. Bootstrap analyses of real data further revealed intervention specific model effects on attention, sentiment and polarization.
\end{abstract}


\section{Introduction}
During an escalating online public opinion event, authorities must select interventions whose timing, content and intensity may determine whether the event subsides or intensifies~\cite{zareer2026opiniondynamics}. This decision is counterfactual because the relevant question is how the event would evolve under a chosen action rather than how it will evolve. In Pearl’s causal hierarchy, the former is associational, whereas the latter occupies the third rung~\cite{pearl2018bookofwhy}. Existing point process and cascade representation methods remain associational~\cite{sun2023castformer,xu2025popularity}. Historical conditional comparisons are confounded because intervention assignment depends on event state. Official announcements, for example, often occur near peak attention, so subsequent natural decline may be misattributed to the intervention.

Recent counterfactual methods either target structured treatments in healthcare~\cite{bica2020crn,hess2024gtransformer} or estimate the effect of a single action~\cite{tian2026citrus}, leaving the broader intervention space unaddressed. Simulation based approaches can generate counterfactual scenarios, but their conclusions depend on behavioral assumptions and are difficult to calibrate with real data~\cite{zareer2026opiniondynamics,cheng2025llmsim,park2023generative}. Counterfactual trajectories should therefore be learned from real propagation cascades with intervention annotations under an explicit causal model~\cite{scholkopf2021causalrep}.

A counterfactual public opinion projection framework is proposed for intervention decision making in this paper. Governance interventions are modeled as exogenous variables, integrating causal structure discovery, counterfactual trajectory generation and intervention strategy recommendation. Two design choices distinguish this framework from prior work. Each intervention is represented by its timing, type, content and intensity rather than a binary indicator. Moreover, observed data reveal who reposted whom but not who triggered whom, while counterfactual outcomes remain unobserved. Ground truth evaluation of causal structure is therefore conducted mainly in semisynthetic settings. Real data are used for structural evaluation on a manually annotated subset, factual trajectory forecasting and model based effect estimation. The main contributions are as follows.
\begin{itemize}
	\item A causal discovery module compresses the candidate edge space and uses a structure aware graph neural network (GNN) to validate explicit edges and discover implicit edges. Structure recovery is evaluated on semisynthetic data with known causal ground truth and a manually annotated real data subset.
	\item A counterfactual generation module extends adversarial balancing from low dimensional structured treatments in healthcare to high dimensional semantic interventions defined by type, content, timing and intensity, enabling trajectory level generation.
	\item An event-level Weibo dataset with intervention annotations is built. It supports interval estimation of model-based counterfactual effects and strategy recommendation for four interventions, uncovering heterogeneous, uncertain effects on attention, sentiment and polarization.
\end{itemize}

The remainder of this paper is organized as follows. Section~\ref{Sec2} reviews related work. Section~\ref{SecCausal} introduces the three rungs of Pearl's causal hierarchy that underpin this study. Section~\ref{Sec3} presents the system model and problem formulation. Section~\ref{SecData} describes dataset construction. Section~\ref{Sec4} details the three modules. Section~\ref{Sec5} reports the experiments and discussion. Section~\ref{Sec6} concludes the paper.

\section{Related Work}\label{Sec2}

The task comprises causal structure discovery, counterfactual trajectory generation and intervention strategy recommendation. Four related research streams are reviewed.

\subsection{Information Cascade Forecasting}

Information cascade forecasting predicts future diffusion from observed histories and follows two main approaches. Point process methods model diffusion as a self exciting event stream in which each event temporarily increases the intensity of subsequent events. Early studies used statistical features and threshold rules~\cite{cheng2014cascades}. Hawkes processes and their neural variants use excitation kernels to quantify the influence of historical events on current intensity~\cite{mei2017neuralhawkes}. SEISMIC fits tweet popularity with a self exciting point process~\cite{zhao2015seismic}, whereas reinforced Poisson processes introduce memory effects to model popularity decay~\cite{shen2014reinforced}.

Representation learning methods instead encode cascades as vectors and regress their future scale. The encoded information has expanded from sequences to structures and multimodal signals. DeepCas learns cascade representations from random walks~\cite{li2017deepcas}. DeepHawkes applies graph convolution to propagation trees for prediction and interpretation~\cite{cao2017deephawkes}. Cascade convolution and Castformer introduce convolutional sequence modeling and attention mechanisms~\cite{chen2019cascn,sun2023castformer}. Coupled GNN integrates multiple structural views~\cite{cao2020coupledgnn}, while recent work incorporates multimodal features and retrieval augmentation~\cite{xu2025popularity}. However, these methods predict factual evolution from observed histories and cannot estimate outcomes under hypothetical interventions.

\subsection{Causal Structure Discovery}

Counterfactual reasoning requires knowledge of the causal mechanisms through which interventions affect diffusion. Causal structure discovery recovers directed dependencies among variables from observational data. NOTEARS~\cite{zheng2018notears} converts DAG learning from discrete structure search into continuous optimization through an exact differentiable acyclicity constraint. DAGMA~\cite{bello2022dagma} further characterizes acyclicity using M matrices and the log determinant, improving optimization and computational efficiency. However, these methods typically learn a global DAG from repeated observations of fixed variables. In event level cascades, nodes vary across sequences and explicit propagation edges may be missing or noisy. Direct application is therefore difficult. In this paper, the candidate edge space is compressed under temporal constraints, while semantic, temporal and structural information is integrated to validate explicit edges and discover latent triggers. The resulting structure supports subsequent counterfactual public opinion projection.

\subsection{Counterfactual Outcome Estimation}

Causal structures describe dependencies among variables, whereas counterfactual outcome estimation quantifies potential outcomes under alternative interventions. The field originated in clinical decision making, where treatments are assigned over time as patient conditions evolve and outcomes under alternative treatment sequences must be estimated. Existing methods mainly use weighting adjustment, representation balancing or sequence modeling. RMSN~\cite{lim2018rmsn} adjusts for time varying confounding with marginal structural networks. CRN~\cite{bica2020crn} uses a gradient reversal layer (GRL) to adversarially learn treatment invariant balanced representations. G-Net~\cite{li2021gnet} implements g computation with neural networks. Causal Transformer~\cite{melnychuk2022causaltransformer} estimates outcomes under multiple treatments using a Transformer backbone. TECDE~\cite{seedat2022tecde} introduces continuous time modeling, while G-Transformer~\cite{hess2024gtransformer} targets conditional average potential outcomes. Under other treatment settings, DESCN~\cite{zhong2022descn} estimates individual treatment effects for static interventions and marketing ranking, while gradient interpolation extends effect estimation to continuous treatments~\cite{nagalapatti2024ctee}. However, these methods primarily address structured treatments such as drugs, doses and timing, together with single or low dimensional outcomes. Semantic interventions jointly defined by type, content, timing and intensity and multidimensional public opinion trajectories remain underexplored. Representation balancing is therefore adopted to generate intervention conditioned trajectories of attention, sentiment and polarization in this paper.

\subsection{Counterfactual Analysis and Intervention Optimization on Social Media}

Counterfactual methods on social media mainly estimate the effects of misinformation exposure or fact checking. CNTPP~\cite{zhang2022cntpp} uses a counterfactual neural temporal point process to estimate the causal effect of misinformation on social engagement. CITRUS~\cite{tian2026citrus} demonstrates the need for causal modeling in online influence estimation through counterfactual analysis of misinformation exposure. Slaughter et al.~\cite{slaughter2025pnas} apply synthetic control to evaluate the effect of Community Notes on false information diffusion. However, these studies focus on a single exposure or verification action rather than multiple governance interventions.

For intervention optimization, Zhu et al.~\cite{zhu2025structureaware} study node level, time varying and resource constrained rumor control. Other studies use LLM based multiagent or generative agents to simulate public opinion dynamics and intervention effects~\cite{cheng2025llmsim,park2023generative}. The former focuses on resource allocation across network nodes, whereas the latter relies on predefined agent behavior and rarely learns counterfactual trajectories for multiple governance actions from real propagation data with intervention annotations.

Overall, prior work separately addresses cascade forecasting, causal structure discovery, counterfactual estimation and intervention optimization. A unified framework that learns event level structures from real cascades, generates trajectories under multiple interventions and supports strategy recommendation remains lacking.

\section{Causal Hierarchy and Task Formulation}
\label{SecCausal}

\subsection{Illustrative Model}
\label{SecCausalExample}

To illustrate the associational, interventional and counterfactual rungs of Pearl's causal hierarchy~\cite{pearl2018bookofwhy}, consider the decline in attention $\Delta h_i$ for event $i$ over a fixed prediction window
\begin{equation}
	\Delta h_i=\alpha b_i+\beta a_i,
	\label{eq:illustrative_model}
\end{equation}
where $b_i$ denotes the intrinsic evolution baseline and $\alpha$ is the natural decay coefficient. The variable $a_i\in\{0,1\}$ indicates whether an official announcement is issued, while $\beta$ denotes its additional reduction. Thus, $\alpha b_i$ captures the natural decline that may occur regardless of intervention, whereas $\beta a_i$ captures the added intervention effect. The three causal levels differ in whether these components can be correctly separated.

\textbf{First rung: Association.} The associational level compares the mean decline in attention between historical events with and without official announcements
\begin{align}
	&\mathbb{E}[\Delta h\mid a=1]
	-
	\mathbb{E}[\Delta h\mid a=0]
	\nonumber\\
	={}&
	\beta+
	\alpha\left(
	\mathbb{E}[b\mid a=1]
	-
	\mathbb{E}[b\mid a=0]
	\right).
	\label{eq:association_bias}
\end{align}
Official announcements are more likely to be issued when attention or risk is high or near its peak. The evolution baselines of the two groups therefore usually differ
$\mathbb{E}[b\mid a=1]\neq\mathbb{E}[b\mid a=0]$.
Consequently, the difference between groups combines the announcement effect $\beta$ with natural decline caused by baseline differences. Attributing the entire difference to the announcement introduces selection bias.

\textbf{Second rung: Intervention.} The interventional level places the same target population under two intervention settings. Only the intervention $a$ is changed, while the distribution of event baselines remains fixed
\begin{equation}
	P(b\mid\mathrm{do}(a=1))
	=
	P(b\mid\mathrm{do}(a=0))
	=
	P(b).
\end{equation}
The difference in mean attention decline between the two settings is therefore
\begin{align}
	&\mathbb{E}[\Delta h\mid\mathrm{do}(a=1)]
	-
	\mathbb{E}[\Delta h\mid\mathrm{do}(a=0)]
	\nonumber\\
	={}&
	\left(\alpha\mathbb{E}[b]+\beta\right)
	-
	\alpha\mathbb{E}[b]
	=\beta.
	\label{eq:intervention_effect}
\end{align}
Both hypothetical settings concern the same event population and thus share the mean baseline $\mathbb{E}[b]$. The natural decline terms cancel, leaving $\beta$ as the population average causal effect of the announcement.

\textbf{Third rung: Counterfactuals.} The interventional level estimates the average effect of an announcement on an event population. The counterfactual level instead asks how a specific event would have evolved under a different action while holding its baseline $b_i$ fixed.

Suppose event $i$ received an official announcement with $a_i=1$. The observed decline in attention is
\begin{equation}
	\Delta h_i(1)=\alpha b_i+\beta.
	\label{eq:factual_outcome}
\end{equation}
Post hoc counterfactual inference first recovers the event baseline from the factual outcome
\begin{equation}
	b_i=\frac{\Delta h_i(1)-\beta}{\alpha}.
	\label{eq:abduction_base}
\end{equation}
The baseline $b_i$ is then held fixed while the intervention is changed from $a_i=1$ to $a_i=0$, yielding the potential outcome without the announcement
\begin{equation}
	\Delta h_i(0)=\alpha b_i.
	\label{eq:counterfactual_outcome}
\end{equation}
The counterfactual effect of the announcement on this event is therefore
\begin{equation}
	\Delta h_i(1)-\Delta h_i(0)=\beta.
	\label{eq:individual_effect}
\end{equation}

This procedure follows abduction, action and prediction. The event baseline is inferred from the factual outcome, the intervention is replaced while the baseline remains fixed, and the alternative outcome is then predicted.

In summary, association compares historical groups with potentially different baselines and is therefore vulnerable to selection bias. Intervention holds the baseline distribution constant to estimate a population average effect. Counterfactual inference fixes the baseline of a specific event and changes only the intervention to estimate its alternative outcome.

\subsection{Mapping to Counterfactual Public Opinion Projection}
\label{SecCausalMapping}

Real public opinion dynamics are neither one dimensional nor linear. Eq.~\eqref{eq:illustrative_model} only separates natural evolution from intervention effects to illustrate the causal hierarchy. The scalar baseline, binary intervention and single attention outcome are extended to an event state representation, a semantic intervention representation and a multidimensional public opinion trajectory. Their relationships are modeled by a nonlinear generative model. The correspondence is shown in Table~\ref{tab:toy_mapping}.

The basic counterfactual logic is retained. An event state $z$ is first extracted from observed information. The state is then fixed while the intervention representation $a_{\mathrm{repr}}$ is replaced. The decoder is rerun to generate the potential trajectory under the candidate intervention. At the population level, adversarial balancing reduces differences between state distributions across intervention groups. At the event level, fixing the state and replacing the intervention enables event level counterfactual projection. Formal definitions are provided below.

\begin{table}[!t]
	\centering
	\caption{Mapping between the illustrative model and Counterfactual Public Opinion Projection}
	\label{tab:toy_mapping}
	\renewcommand{\arraystretch}{1.2}
	\small
	\begin{tabular}{
			p{0.33\columnwidth}
			p{0.56\columnwidth}}
		\toprule
		\textbf{Illustrative model}
		&
		\textbf{Corresponding variable or equation}
		\\
		\midrule
		
		Event baseline $b_i$
		&
		Event state $z$ in Eq.~\eqref{eq:sharedz}
		\\
		
		Natural decline $\alpha b_i$
		&
		Decoder output without intervention in Eq.~\eqref{eq:decoder}
		\\
		
		Binary intervention $a_i$
		&
		Intervention representation $a_{\mathrm{repr}}$ in Eq.~\eqref{eq:intervenc}
		\\
		
		Announcement effect $\beta a_i$
		&
		Decoder outputs under different interventions in Eq.~\eqref{eq:decoder}
		\\
		
		Attention decline $\Delta h_i$
		&
		Counterfactual outcome $\hat{Y}(a)$ in Eq.~\eqref{eq:rq2target}
		\\
		
		$\Delta h_i=\alpha b_i+\beta a_i$
		&
		Counterfactual trajectory decoder in Eqs.~\eqref{eq:decoder} and~\eqref{eq:sampling}
		\\
		
		Balanced baseline distributions
		&
		Adversarial balancing in Eq.~\eqref{eq:grl}
		\\
		
		Fixed $b_i$ with replaced $a_i$
		&
		Fixed $z$ with replaced $a_{\mathrm{repr}}$ in Eq.~\eqref{eq:decoder}
		\\
		\bottomrule
	\end{tabular}
\end{table}

\section{System Model and Problem Formulation}\label{Sec3}

\subsection{Event Level Propagation Sequence}

For a public opinion event $\mathcal{E}$, its event level propagation sequence within the observation window $[0,T_{obs}]$ is defined as
\begin{equation}
	\mathcal{S} = \{e_1, e_2, \ldots, e_n\}, \quad e_i = (t_i, u_i, c_i, \tau_i),
	\label{eq:sequence}
\end{equation}
where $t_i$, $u_i$, and $c_i$ denote the timestamp, author, and textual content. The event type is denoted by $\tau_i \in \{\text{original post},\text{repost},\text{comment}\}$.

Causal discovery aims to identify which event triggered another. Each original post, repost, or comment is therefore treated as an individual node. User level aggregation would merge multiple posts from the same user and compress causal chains, whereas aggregation into fixed time windows would obscure specific triggering relations. The sequence is ordered chronologically as $t_1 \le t_2 \le \cdots \le t_n$, so the node index $i$ naturally preserves temporal order.

\subsection{Governance Intervention}\label{Sec3B2}

A governance intervention is defined as
\begin{equation}
	a = (t^*, \text{type}, \text{content}, \text{intensity}),
	\label{eq:intervention}
\end{equation}
where the four components denote intervention timing, type, content, and intensity. The sequence $\mathcal{S}$ describes natural propagation without external intervention, whereas $a$ is actively specified in counterfactual public opinion projection. These four dimensions jointly determine the causal effect. For example, the same announcement may produce markedly different effects when released two hours earlier or by a provincial rather than municipal account.

\subsection{Causal Graph}\label{Sec3C}

Causal triggering relations between propagation events form a directed graph
\begin{equation}
	G = (\mathcal{S}, \mathcal{A}), \quad \mathcal{A} \subseteq \mathcal{S} \times \mathcal{S},
	\label{eq:causalgraph}
\end{equation}
where $\mathcal{S} \times \mathcal{S}$ contains all ordered pairs $(e_i,e_j)$, and $\mathcal{A}$ is the subset representing causal triggers. An edge $e_i \to e_j$ may exist only when $t_i<t_j$ and indicates that the publication of $e_i$ causally triggered $e_j$, rather than merely preceding it or sharing similar content. Because every edge satisfies this temporal constraint, $G$ is a directed acyclic graph by construction, and the node indices define a valid topological order.

\subsection{Operational Definition of Macroscopic Trajectories}\label{Sec3D}

Counterfactual generation and intervention recommendation require quantitative representations of event states. Because raw propagation data comprise discrete events, three macroscopic measures are aggregated over fixed windows to produce regularly spaced time series for sequence modelling and cross event comparison.

\textbf{Attention} measures propagation volume and is defined as the weighted activity within window $t$
\begin{equation}
	h(t) = \sum_{e_i \in \mathcal{W}(t)} w_{\tau_i}, \quad
	w_{\tau} = \{w_{\text{original}},w_{\text{repost}},w_{\text{comment}}\},
	\label{eq:heat}
\end{equation}
where $\mathcal{W}(t)$ is the event set within window $t$, and $w_{\tau}$ is the corresponding type weight. Original posts and reposts receive greater weights because they further disseminate content, whereas comments mainly respond to existing posts.

\textbf{Sentiment} measures the prevailing orientation and is defined as the mean sentiment score within window $t$
\begin{equation}
	s(t) = \frac{1}{|\mathcal{W}(t)|}
	\sum_{e_i \in \mathcal{W}(t)} \sigma(c_i),
	\quad \sigma(\cdot) \in [-1,1],
	\label{eq:sentiment}
\end{equation}
where $\sigma(c_i)$ is the sentiment score of content $c_i$.

\textbf{Polarization} measures the divergence of opinions within a window. Given the absence of a universal definition, it is operationalized using sentiment variance
\begin{equation}
	p(t) = 4 \cdot
	\mathrm{Var}\!\left[
	\frac{\sigma(c_i)+1}{2}
	\right]_{e_i \in \mathcal{W}(t)}.
	\label{eq:polarization}
\end{equation}
Sentiment scores are first mapped from $[-1,1]$ to $[0,1]$, after which their variance is multiplied by four. This yields $p(t)\in[0,1]$. Polarization approaches zero when opinions are homogeneous and one when they are evenly split between the two extremes.

\textbf{Potential outcome} under intervention $a$ is defined from these three measures as
\begin{equation}
	Y(a)
	=
	\big(
	h(t;a),\,
	s(t;a),\,
	p(t;a)
	\big)_{t>T_{\mathrm{obs}}},
	\label{eq:potential}
\end{equation}
where $h(t;a)$, $s(t;a)$, and $p(t;a)$ denote attention, sentiment, and polarization at time $t$ under intervention $a$. Together, they form the post intervention trajectory. For any event, only the outcome $Y(a^{\mathrm{obs}})$ under the observed intervention can be observed. Potential outcomes $Y(a')$ under alternative interventions $a'\neq a^{\mathrm{obs}}$ remain unobserved, which constitutes the fundamental problem of causal inference.

\subsection{Formal Objectives of the Three Research Questions}\label{Sec3E}

\textbf{Question 1: Causal structure discovery.}
Estimate the causal edge set $\hat{\mathcal{A}}$ from $\mathcal{S}$.

\textbf{Question 2: Counterfactual trajectory generation.}
Given the causal graph $G$ and a candidate intervention $a$, estimate the potential outcome defined in Eq.~\eqref{eq:potential}
\begin{equation}
	\hat{Y}(a)
	=
	\big(
	\hat{h}(t;a),\,
	\hat{s}(t;a),\,
	\hat{p}(t;a)
	\big)_{t>T_{\mathrm{obs}}}.
	\label{eq:rq2target}
\end{equation}
Representative simulated content
$\hat{\mathcal{C}}(a)=\{\hat{c}_1,\ldots,\hat{c}_m\}$
may also be generated to illustrate the candidate scenario. This content only supplements the macroscopic trajectory and is excluded from potential outcome estimation, effect calculation, strategy ranking, and experimental evaluation.

\textbf{Question 3: Intervention recommendation.}
Given a candidate intervention set
$\mathcal{A}_{cand}=\{a_1,\ldots,a_K\}$,
perform Counterfactual Public Opinion Projection for each $a_k$, compare their multiobjective effects, and return the Pareto frontier with causal pathway explanations.

The recommendation module evaluates only the candidate set specified by the decision maker. It does not enumerate the full combinatorial space of intervention tuples.

\begin{figure*}[!t]
	\centering
	\includegraphics[width=0.9\textwidth]{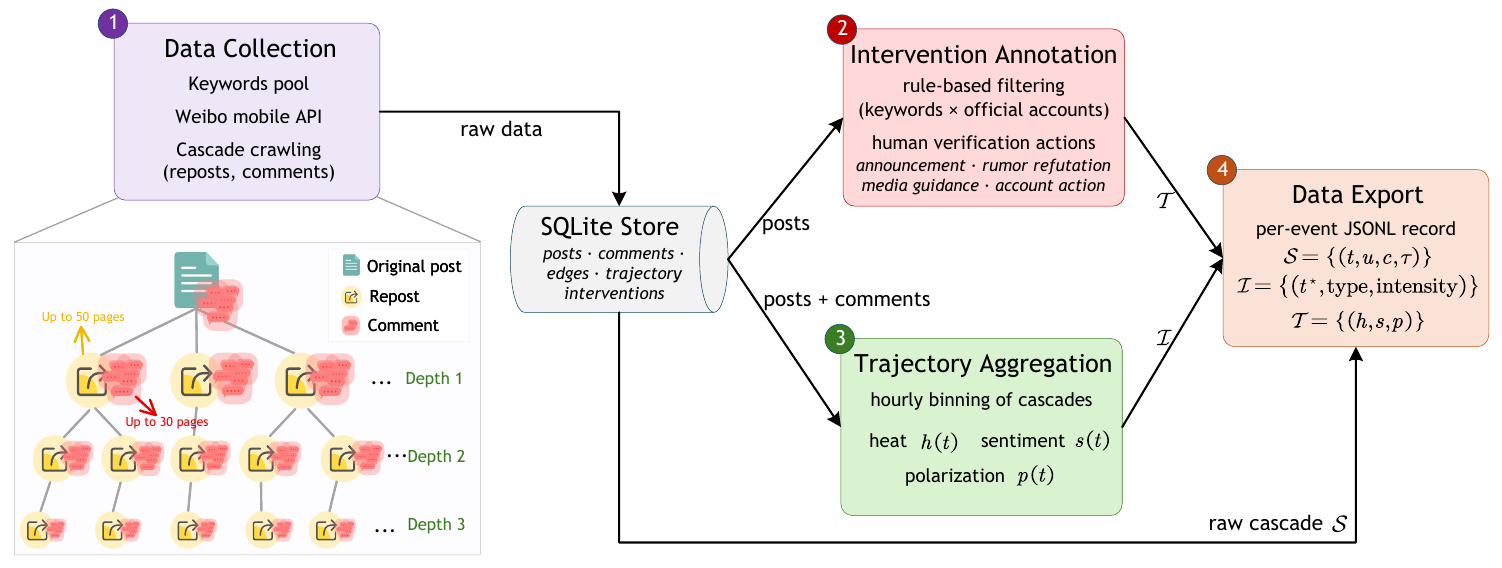}
	\caption{Data collection and processing pipeline.}
	\label{fig:pipeline}
\end{figure*}

\section{Dataset Construction}\label{SecData}

Existing Chinese public opinion datasets mainly record propagation structures and textual content but rarely annotate governance interventions, limiting their use for Counterfactual Public Opinion Projection. A public opinion event dataset with verified intervention annotations was therefore constructed from Sina Weibo at m.weibo.cn
\begin{equation}
	\mathcal{D}
	=
	\{(\mathcal{S}_j,\mathcal{I}_j,\mathcal{T}_j)\}_{j=1}^{N},
	\label{eq:dataset}
\end{equation}
where $\mathcal{S}_j$ is the propagation sequence of event $j$ defined in Eq.~\eqref{eq:sequence}, and $\mathcal{I}_j$ is its verified intervention set with elements defined by Eq.~\eqref{eq:intervention}. When no governance intervention occurs, $\mathcal{I}_j=\varnothing$. The macroscopic trajectory $\mathcal{T}_j$ comprises attention, sentiment, and polarization. Dataset construction consists of cascade collection, intervention annotation, trajectory aggregation, and data export, as shown in Fig.~\ref{fig:pipeline}.

\subsection{Cascade Collection and Storage}

32 seed queries covering official announcements, rumor rebuttals, media guidance, and account moderation are used to identify candidate events. For each retrieved original post, repost relations and comments are collected recursively. Repost trees are limited to a depth of three and 50 pages per node, while comments are limited to 30 pages. Each resulting cascade comprises the original post, reposts, and comments. Only publicly accessible content is collected. User identifiers are irreversibly hashed before storage, and no plaintext identity information is retained.

\subsection{Intervention Annotation and Scenario Partitioning}

Governance actions rather than individual posts are annotated. Repeated releases or descriptions of the same action are merged into one record, with the earliest occurrence recorded as $t^*$. Annotation follows a three stage filtering process
\begin{equation}
	\mathcal{P}
	\xrightarrow{\text{rule screening}}
	\mathcal{C}_1
	\xrightarrow{\text{type classification}}
	\mathcal{C}_2
	\xrightarrow{\text{manual verification}}
	\mathcal{I},
	\label{eq:annotation}
\end{equation}
where $\mathcal{P}$ contains all posts in a cascade. Rule screening retrieved candidates using intervention keywords and account verification information. Candidates are then assigned to four types according to keyword priority and action target. Manual verification determines whether each candidate constitutes a governance intervention and whether its type is correct.

The verified events are partitioned into natural evolution and intervention sets
\begin{align}
	\mathcal{D}_{nat}
	&=
	\{j\in\{1,\ldots,N\}:\mathcal{I}_j=\emptyset\},
	\nonumber\\
	\mathcal{D}_{iv}
	&=
	\{j\in\{1,\ldots,N\}:\mathcal{I}_j\neq\emptyset\}.
	\label{eq:observed_scenarios}
\end{align}
For an event with multiple verified interventions, the earliest action is treated as the observed intervention
\begin{equation}
	a_j^{\mathrm{obs}}
	=
	\arg\min_{a\in\mathcal{I}_j} t^*(a).
	\label{eq:factual_intervention}
\end{equation}
The counterfactual query set is then constructed as
\begin{equation}
	\mathcal{D}_{cf}
	=
	\left\{
	(j,a'):
	j\in\mathcal{D}_{iv},\
	a'\in
	\big(\mathcal{A}_{cand}\cup\{a_0\}\big)
	\setminus\{a_j^{\mathrm{obs}}\}
	\right\},
	\label{eq:scenarios}
\end{equation}
where $a_0$ denotes no intervention and $a'$ is a candidate action distinct from the observed intervention. The set $\mathcal{D}_{cf}$ contains event and candidate action pairs for counterfactual public opinion projection rather than additional observations.

\subsection{Trajectory Aggregation and Export}

Following Eqs.~\eqref{eq:heat}--\eqref{eq:polarization}, each propagation cascade is aggregated over consecutive windows of one hour to obtain
\begin{equation}
	\mathcal{T}_j
	=
	\big(h_j(t),s_j(t),p_j(t)\big)_{t=1}^{T_j}.
\end{equation}
The attention weights are set to
$w_{\text{original}}=w_{\text{repost}}=1$
and
$w_{\text{comment}}=0.5$.
For each text $c_i$, SnowNLP produces a positive sentiment probability $q(c_i)\in[0,1]$, which is mapped to $[-1,1]$ using
$\sigma(c_i)=2q(c_i)-1$.
The propagation sequence $\mathcal{S}_j$, explicit edge set $\mathcal{E}_{\mathrm{explicit}}^{(j)}$, intervention annotations $\mathcal{I}_j$, and macroscopic trajectory $\mathcal{T}_j$ are exported for each event.

\subsection{Dataset Scale and Evaluation Protocol}\label{SecDataScale}

Among 621 candidate events, 539 yield valid propagation cascades and macroscopic trajectories. The remainder are excluded because repost relations could not be collected or too few valid time windows are available. Verified governance interventions are identified in 120 events, most commonly official announcements and media guidance. Publicly retrievable rumor rebuttals and account moderation actions that could be matched to complete cascades are less frequent. Dataset statistics are reported in Table~\ref{tab:dataset}.

\begin{table}[!t]
	\centering
	\caption{Dataset statistics}
	\label{tab:dataset}
	\begin{tabular}{lr}
		\toprule
		Metric & Value \\
		\midrule
		Seed queries & 32 \\
		Candidate events collected & 621 \\
		Valid propagation cascades & 539 \\
		Total cascade nodes & 104{,}444 \\
		Explicit repost edges & 116{,}906 \\
		Intervention candidates from rule screening & 162 \\
		Events with verified interventions & 120 \\
		\quad Official announcement
		& 50 \\
		\quad Media guidance
		& 47 \\
		\quad Rumor rebuttal
		& 12 \\
		\quad Account moderation
		& 11 \\
		Total macroscopic trajectory windows & 10{,}843 \\
		\bottomrule
	\end{tabular}
\end{table}

All valid cascades are used for causal structure discovery, while events satisfying the observation and prediction window requirements are used to train the counterfactual generation model. Because counterfactual outcomes are unobservable for real events, structure recovery is evaluated on semisynthetic data with known causal ground truth. Real data are used to evaluate factual trajectory prediction and report model estimated intervention effects and case results.

\textbf{Semisynthetic data.}
A cascade simulator with known causal ground truth is constructed to evaluate structure recovery. To prevent the task from reducing to semantic similarity matching, nodes are generated through three mechanisms. Content similar causal children comprises 60\% of the nodes and are obtained by adding noise to the parent representation. Content dissimilar causal children comprises 25\%, with independently generated content but timestamps shortly after their parents. The remaining 15\% are exogenous events without parents. Cascade sizes are sampled with replacement from the empirical distribution of Weibo16~\cite{cao2017deephawkes}. To simulate incomplete platform observations, only 70\% of true causal edges are exposed, while spurious explicit edges are assigned to 10\% of exogenous events. These settings evaluate latent edge discovery and explicit edge validation, respectively.

\section{Proposed Method}\label{Sec4}

\begin{figure*}[t]
	\centering
	\includegraphics[width=0.9\textwidth]{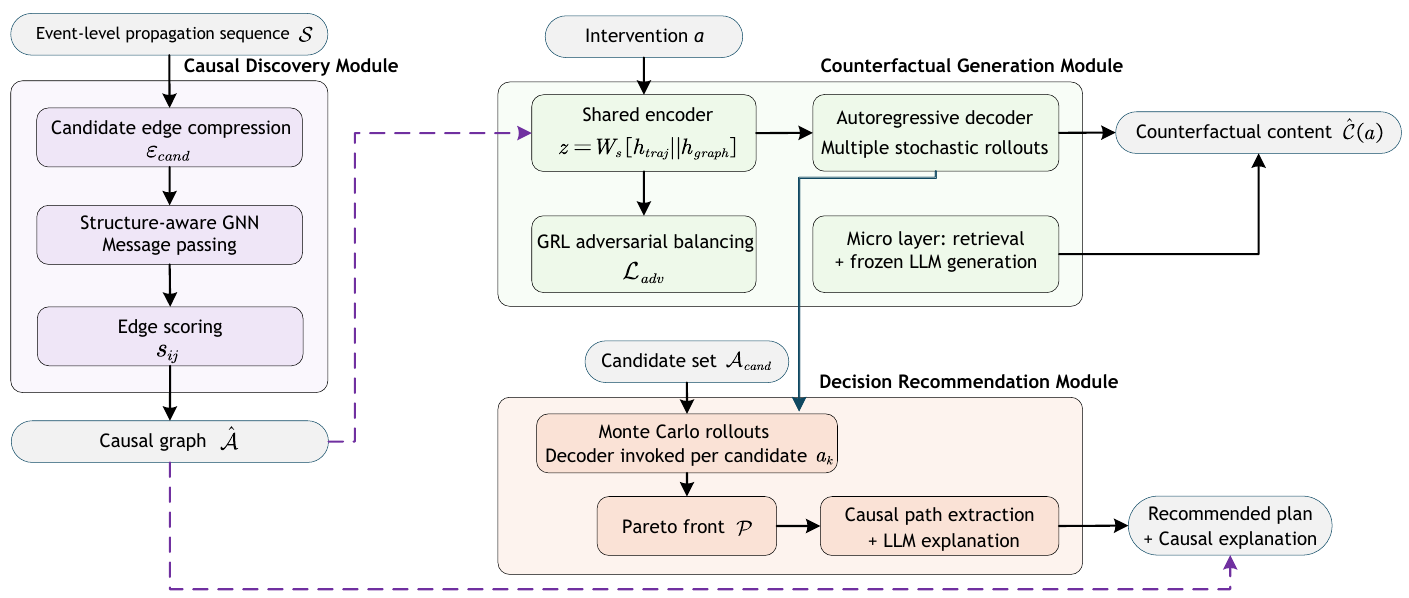}
	\caption{Overview of the proposed framework.}
	\label{fig:architecture}
\end{figure*}

This section presents the causal discovery, counterfactual generation, and intervention recommendation modules corresponding to the three research questions in Section \ref{Sec3E}. Causal discovery provides structural summaries for counterfactual generation and propagation paths for recommendation explanations. Counterfactual generation evaluates candidate interventions through repeated stochastic rollouts. The overall framework is shown in Fig.~\ref{fig:architecture}.

\subsection{Causal Discovery Module}\label{Sec4A}

This module addresses Question 1 by estimating causal triggering edges from a propagation sequence and constructing
$\hat{G}=(\mathcal{S},\hat{\mathcal{A}})$.
Enumerating all ordered pairs in a cascade of $n$ nodes requires $O(n^2)$ computation. The search space is reduced by combining explicit platform edges with nonexplicit node pairs satisfying temporal and semantic constraints
\begin{multline}
	\mathcal{E}_{cand}
	=
	\mathcal{E}_{explicit}
	\cup
	\{(e_i,e_j)\mid
	t_i<t_j\leq t_i+w,\\[-1pt]
	\mathrm{sim}(c_i,c_j)>\theta
	\},
	\label{eq:candidate}
\end{multline}
where $w$ is the temporal window and $\theta$ is the semantic similarity threshold. Explicit edges are oriented along the propagation direction during preprocessing, and only relations satisfying $t_i<t_j$ are retained. If each node has an average of $k$ candidate neighbours, the candidate set has size $O(nk)$, where $k\ll n$. Latent edge discovery is thus restricted to temporally proximate and semantically related node pairs.

To prevent unverified candidate edges from distorting node representations, message passing is performed only over the explicit propagation graph. Each node input $x_i$ concatenates its text embedding and temporal features and is projected as
\begin{equation}
	h_i^{(0)}=W_{in}x_i.
	\label{eq:gnninit}
\end{equation}
A graph attention network then aggregates the explicit neighbourhood
\begin{equation}
	h_i^{(l+1)}
	=
	\sigma\left(
	\sum_{j\in N_{explicit}(i)\cup\{i\}}
	\alpha_{ij}^{(l)}W^{(l)}h_j^{(l)}
	\right),
	\label{eq:gnnupdate}
\end{equation}
where $\alpha_{ij}^{(l)}$ is the normalized attention coefficient at layer $l$, and $N_{explicit}(i)$ is the explicit neighbourhood of node $e_i$. A total of $L$ layers are stacked.

For each candidate edge $(e_i,e_j)$, the endpoint representations, normalized time difference, and explicit graph distance are passed to a multilayer perceptron
\begin{equation}
	s_{ij}
	=
	\mathrm{MLP}
	\left(
	\left[
	h_i^{(L)}
	\|
	h_j^{(L)}
	\|
	\phi(\Delta t_{ij})
	\|
	\psi(d_{tree}(i,j))
	\right]
	\right),
	\label{eq:edgescore}
\end{equation}
where $\phi(\Delta t_{ij})$ is the normalized representation of
$\Delta t_{ij}=t_j-t_i$ and captures temporal proximity. Shorter intervals generally indicate stronger triggering relations. The feature $\psi(d_{tree}(i,j))$ represents the hop distance from both endpoints to their lowest common ancestor in the explicit tree. It supplements the limited receptive field of $L$ message passing layers and is computed independently of message passing.

\textbf{Training and loss functions.}
The causal discovery module is jointly optimized for sparsity, semantic reconstruction, and explicit edge preservation. The predicted probability of a candidate edge $(e_i,e_j)$ is
\begin{equation}
	p_{ij}=\operatorname{sigmoid}(s_{ij}).
\end{equation}
Given a decision threshold $\delta$, the estimated edge set is
\begin{equation}
	\hat{\mathcal{A}}
	=
	\{(e_i,e_j)\in\mathcal{E}_{cand}\mid p_{ij}\geq\delta\}.
	\label{eq:estimated_edges}
\end{equation}
The total loss is
\begin{equation}
	\mathcal{L}_{causal}
	=
	\lambda_1\mathcal{L}_{sparse}
	+
	\lambda_2\mathcal{L}_{fid}
	+
	\lambda_3\mathcal{L}_{prior}.
	\label{eq:losscausal}
\end{equation}

The sparsity loss is the mean candidate edge probability
\begin{equation}
	\mathcal{L}_{sparse}
	=
	\frac{1}{|\mathcal{E}_{cand}|}
	\sum_{(i,j)\in\mathcal{E}_{cand}}p_{ij}.
\end{equation}
This term penalizes candidate edge probabilities and discourages dense graphs by retaining only edges that support semantic reconstruction. For each node $e_j$, candidate parent representations are aggregated according to their edge probabilities
\begin{equation}
	\bar{h}_j
	=
	\frac{
		\sum_{i:(i,j)\in\mathcal{E}_{cand}}
		p_{ij}h_i^{(L)}
	}{
		\sum_{i:(i,j)\in\mathcal{E}_{cand}}p_{ij}+\epsilon
	}.
\end{equation}
Here, $\epsilon$ is a small constant that prevents division by zero. The aggregated representation is projected into the text embedding space
\begin{equation}
	\hat{x}^{\,text}_j=W_r\bar{h}_j,
\end{equation}
where $W_r$ is a learnable reconstruction matrix. Let $\mathcal{V}_{in}$ denote nodes with at least one candidate parent. The reconstruction loss is
\begin{equation}
	\mathcal{L}_{fid}
	=
	\frac{1}{|\mathcal{V}_{in}|}
	\sum_{j\in\mathcal{V}_{in}}
	\left\|
	\hat{x}^{\,text}_j-x^{text}_j
	\right\|_2^2,
\end{equation}
where $x^{text}_j$ is the original text embedding of node $e_j$. This loss requires the weighted parent representation to reconstruct the semantics of its child. The explicit edge prior is
\begin{equation}
	\mathcal{L}_{prior}
	=
	-\frac{1}{|\mathcal{E}_{explicit}|}
	\sum_{(i,j)\in\mathcal{E}_{explicit}}
	\log p_{ij}.
\end{equation}
This term softly favours explicit platform edges while allowing their probabilities to decrease when structural and semantic evidence is insufficient.

Because every candidate edge satisfies $t_i<t_j$, any candidate subgraph is acyclic and requires no additional acyclicity constraint. After training, the module is frozen, and its node representation summary and predicted edges are passed to subsequent modules.

\subsection{Counterfactual Generation Module}\label{Sec4B}

This module addresses Question 2. It contains a macroscopic layer that generates trajectories $\hat{Y}(a)$ and a microscopic layer that generates representative simulated content $\hat{\mathcal{C}}(a)$.

\textbf{Macroscopic layer.}
Governance interventions are not randomly assigned. Official announcements, for example, are often released when attention is high or approaching its peak. Differences between intervention and control trajectories therefore reflect both intervention effects and event state differences. To mitigate this selection bias, an adversarial balancing mechanism based on CNTPP~\cite{zhang2022cntpp} and a gradient reversal layer or GRL~\cite{ganin2016grl} is adopted.

The observed trajectory is denoted by
$y_{\leq T_{\mathrm{obs}}}
=[(h(t),s(t),p(t))]_{t=1}^{T_{\mathrm{obs}}}$.
A GRU encodes the trajectory history, while mean pooling over the node representations from causal discovery produces a structural summary
\begin{equation}
	h_{\mathrm{traj}}
	=
	\mathrm{LayerNorm}
	\big(
	\mathrm{GRU}(y_{\leq T_{\mathrm{obs}}})
	\big),
	\label{eq:trajenc}
\end{equation}
\begin{equation}
	\bar{g}
	=
	\frac{1}{n}\sum_{i=1}^{n}h_i^{(L)},\qquad
	h_{\mathrm{graph}}
	=
	\mathrm{LayerNorm}(W_g\bar{g}).
	\label{eq:graphenc}
\end{equation}
The trajectory and graph summaries are concatenated and projected into an event state representation
\begin{equation}
	z = W_s \big[\, h_{\mathrm{traj}} \,\|\, h_{\mathrm{graph}} \,\big].
	\label{eq:sharedz}
\end{equation}

Although the intervention is not explicitly included in $z$, the observed history may still reveal treatment assignment. A treatment predictor is therefore attached to $z$. Let $r_i\in\{0,1\}$ indicate whether event $i$ received an intervention. The predicted probability and weighted binary cross entropy are
\begin{equation}
	q_i=\operatorname{sigmoid}\!\left(
	\mathrm{MLP}_{\mathrm{adv}}(\mathrm{GRL}_{\lambda}(z_i))
	\right),
\end{equation}
\begin{equation}
	\mathcal{L}_{\mathrm{adv}}
	=-\frac{1}{N}\sum_{i=1}^{N}
	\left[
	\omega_1r_i\log q_i+
	\omega_0(1-r_i)\log(1-q_i)
	\right],
	\label{eq:grl}
\end{equation}
where $\omega_1$ and $\omega_0$ correct the imbalance between treatment and control samples. The layer $\mathrm{GRL}_{\lambda}$ preserves its input during forward propagation but multiplies encoder gradients by $-\lambda$ during backpropagation. The predictor distinguishes treatment status, whereas the state encoder reduces its identifiability from $z$, thereby improving representation balance. This mechanism cannot guarantee the removal of unobserved confounding, and subsequent effects remain model estimates.

Each field of the candidate intervention tuple is encoded separately
\begin{equation}
	\begin{aligned}
		a_{\mathrm{repr}} = \big[\,&\mathrm{Embed}(\text{type})
		\,\|\, \mathrm{Enc}_{\mathrm{text}}(\text{content})\\
		&\|\, t^* \,\|\, \text{intensity}\,\big],
	\end{aligned}
	\label{eq:intervenc}
\end{equation}
where intervention type uses a learnable embedding, content is encoded by a pretrained language model, and timing and intensity are numerical features.

Future trajectories are generated autoregressively from the balanced state $z$ and intervention representation $a_{\mathrm{repr}}$. Following Causal Transformer~\cite{melnychuk2022causaltransformer}, a Transformer decoder predicts the mean and log variance at each step
\begin{equation}
	(\mu_t,\log\sigma_t^2)
	=
	\mathrm{TransformerDecoder}
	\big(z,a_{\mathrm{repr}},y_{<t}\big),
	\label{eq:decoder}
\end{equation}
\begin{equation}
	\hat{y}_t
	\sim
	\mathcal{N}(\mu_t,\sigma_t^2),
	\quad
	t=T_{\mathrm{obs}}+1,\ldots,T_{\mathrm{obs}}+T_{\mathrm{future}}.
	\label{eq:sampling}
\end{equation}
Here, $y_{<t}$ contains all preceding trajectory points. Stochastic autoregressive sampling supports repeated rollouts under the same intervention to estimate its effect distribution. Ground truth preceding points are used during training, whereas generated points are used during inference.

Trajectory and treatment prediction are optimized jointly
\begin{equation}
	\mathcal{L}_{\mathrm{macro}}
	=
	\mathcal{L}_{\mathrm{traj}}
	+
	\mathcal{L}_{\mathrm{adv}}.
	\label{eq:lossmacro}
\end{equation}
The adversarial loss propagates normally through the predictor but reaches the state encoder through the GRL with gradients scaled by $-\lambda$. The trajectory loss is the mean squared error (MSE) between predicted and observed future trajectories
\begin{equation}
	\mathcal{L}_{\mathrm{traj}}
	=
	\frac{1}{T_{\mathrm{future}}}
	\sum_{t=T_{\mathrm{obs}}+1}^{T_{\mathrm{obs}}+T_{\mathrm{future}}}
	\|\mu_t-y_t\|_2^2.
	\label{eq:losstraj}
\end{equation}

\textbf{Microscopic layer.}
The microscopic layer converts numerical trajectories into interpretable scenario descriptions. Historical posts from similar events under similar interventions are retrieved as in context examples~\cite{lewis2020rag}. A frozen LLM then generates representative simulated content $\hat{\mathcal{C}}(a)$ from the event state, candidate intervention, and macroscopic trajectory.

\textbf{Training pipeline.}
The causal discovery module is first trained independently and frozen. Its node representation summary is then used to train the macroscopic counterfactual generation module. Microscopic content generation and intervention recommendation are executed only when required during inference.

\subsection{Intervention Recommendation Module}\label{Sec4C}

This module addresses Question 3 by evaluating and ranking the candidate intervention set $\mathcal{A}_{cand}$ specified by the decision maker. It is an inference only orchestration layer without trainable parameters.

\subsubsection{Pareto Frontier}

For each candidate $a_k\in\mathcal{A}_{cand}$, the macroscopic layer performs $M$ Monte Carlo rollouts and computes the multiobjective trajectory change relative to the no intervention baseline as
$\mathbf{f}(a_k)=(\Delta h,\Delta s,\Delta p)$.
Nondominated sorting over
$\{\mathbf{f}(a_k)\}_{k=1}^{K}$
produces the Pareto frontier
\begin{equation}
	\mathcal{P}
	=
	\{a_k\in\mathcal{A}_{cand}
	\mid
	\nexists\,a_j\in\mathcal{A}_{cand},\,
	a_j\succ a_k\},
	\label{eq:pareto}
\end{equation}
where $a_j\succ a_k$ indicates that $a_j$ is no worse in every objective and strictly better in at least one. Lower attention and polarization and more positive sentiment are preferred.

\subsubsection{Causal Pathway Explanation}

Effect vectors identify preferable interventions but do not explain how the current state emerged. For the current event, the path with the highest cumulative score to a target node is therefore extracted from $\hat{\mathcal{A}}$. Because the graph is acyclic and the node indices define a valid topological order, dynamic programming gives
\begin{equation}
	\mathrm{Best}(v)
	=
	\max_{u\mid(u,v)\in\hat{\mathcal{A}}}
	\left[
	\mathrm{Best}(u)+s_{uv}
	\right],
	\quad v=1,\ldots,n,
	\label{eq:pathdp}
\end{equation}
where $s_{uv}$ is the edge score defined in Eq.~\eqref{eq:edgescore}, and $\mathrm{Best}$ is set to zero for nodes without incoming edges. Backtracking returns the path nodes and edge scores as a structured explanation. A frozen LLM may verbalize this output but does not alter path extraction or intervention ranking. The resulting pathway explains how the current state formed and provides context for decision making.

\section{Experiments}\label{Sec5}

Five experiments were conducted to address the three research questions in Section \ref{Sec3E}.

\subsection{Experimental Setup}\label{Sec5setup}

\textbf{Semisynthetic data.}
The simulator described in Section \ref{SecDataScale} generates 200 training cascades and 20 nonoverlapping cascades held aside for evaluation.

\textbf{Real data.}
The causal discovery module is trained on all 539 cascades. The counterfactual generation module uses 378 events, including 82 treatment and 296 control samples. The observation and prediction windows are set to
$T_{\mathrm{obs}}=12$ hours and
$T_{\mathrm{future}}=24$ hours.
Stratification by treatment status produces 304 training, 37 validation, and 37 test samples.

Structure recovery on real data is evaluated using 45 cascades sampled from the full set. At most three explicit edges and four latent edge candidates are selected from each cascade. AI assisted preannotation followed by two rounds of manual review produces 290 unique edges after deduplication. These comprise 145 causal and 145 noncausal edges, with 135 explicit edges and 155 latent candidates. The annotations are excluded from training, and evaluation is restricted to this subset to avoid treating unlabeled edges as false positives.

\textbf{Implementation details.}
The causal discovery module uses 768 dimensional text representations from text2vec-base-chinese. Candidate edge parameters are set to $w=6$ h and $\theta=0.6$. The GNN containes four layers with 128 hidden units. Adam is used with a learning rate of $10^{-3}$ and batch size 4. Training runs for 80 epochs on semisynthetic data and 30 epochs on real data. The counterfactual generation module uses 128 hidden units and a Transformer decoder with two layers and four attention heads. Adam is used with a learning rate of $3\times10^{-4}$ and batch size 32. Early stopping uses a patience of 30 epochs. The GRL coefficient is gradually increased from 0 to 0.5 to limit disruption of trajectory representations during early training.

\textbf{Evaluation metrics.}
Causal structure recovery is treated as binary classification and evaluated using precision, recall, and F1. An edge is classified as causal when its score exceeds threshold $\tau$. Trajectory prediction is evaluated using MSE between predicted and observed trajectories. Counterfactual effects are computed as the difference between factual intervention and no intervention rollouts. Attention is reported as the cumulative difference over 24 hours, whereas sentiment and polarization are reported as mean differences over the same period. Negative values indicate reductions under intervention.

\begin{table*}[!t]
	\centering
	\caption{Structure recovery on 20 held out semisynthetic cascades}
	\label{tab:rq1syn}
	\begin{tabular}{lccccc}
		\toprule
		Method & Precision & Recall & Average F1 & Similar content recall & Dissimilar content recall \\
		\midrule
		NOTEARS & --- & --- & \textbf{0.8108} & \textbf{0.999} & 0.000 \\
		Proposed causal discovery module & \textbf{0.940} & 0.702 & 0.8056 & 0.689 & \textbf{0.698} \\
		\bottomrule
	\end{tabular}
\end{table*}
\begin{figure}[!t]
	\centering
	\includegraphics[width=0.5\textwidth]{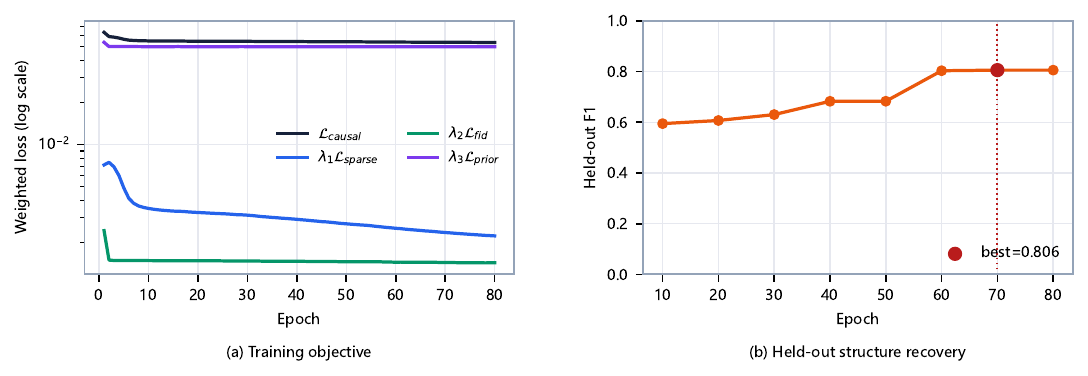}
	\caption{Training dynamics of the causal discovery module.}
	\label{fig:m1_training}
\end{figure}
\subsection{Question 1a. Semisynthetic Structure Recovery}\label{Sec5rq1syn}

NOTEARS~\cite{zheng2018notears} is selected as the baseline. It is fitted independently to each held out cascade using a sparsity coefficient of 0.001, and edges with absolute weights above 0.1 are classified as causal. The proposed module is trained only on the training cascades and then applied directly to the held out cascades. This setting therefore gives NOTEARS greater access to information from each evaluation cascade. The simulator generates 7,439 content similar causal edges and 3,108 content dissimilar causal edges.

Table~\ref{tab:rq1syn} shows similar average F1 values but markedly different performance across edge types. Content similar edges are generated by linear structural equations and therefore matches the assumptions of NOTEARS. For content dissimilar edges, parent and child representations are generated independently and can not be recovered from linear dependence alone. NOTEARS achieves recalls of 0.999 and 0 for the two edge types, compared with 0.689 and 0.698 for the proposed module. Its performance therefore do not arise solely from linear semantic association. The proposed module achieves a precision of 0.940 and recall of 0.702. Recall is also limited by candidate edge coverage of approximately 0.91 because true edges outside the candidate set can not be recovered.

\subsection{Question 1b. Structure Recovery on Real Data}\label{Sec5rq1real}

\begin{table}[!t]
	\centering
	\caption{Structure recovery on 290 labeled real data edges}
	\label{tab:rq1real}
	\setlength{\tabcolsep}{4pt}
	\begin{tabular}{lccccc}
		\toprule
		\multirow{2}{*}{Method} & \multicolumn{3}{c}{Overall} & \multirow{2}{*}{Explicit F1} & \multirow{2}{*}{Latent F1} \\
		\cmidrule(lr){2-4}
		& Precision & Recall & F1 & & \\
		\midrule
		All causal & 0.500 & 1.000 & 0.667 & 0.773 & 0.558 \\
		Heuristic rule & 0.481 & 0.517 & 0.498 & 0.382 & \textbf{0.565} \\
		Proposed $\delta{=}0.5$ & 0.516 & 0.800 & 0.627 & 0.767 & 0.424 \\
		Proposed $\delta{=}0.35$ & \textbf{0.523} & \textbf{0.952} & \textbf{0.675} & \textbf{0.773} & 0.561 \\
		\bottomrule
	\end{tabular}
\end{table}

\begin{figure}[!t]
	\centering
	\includegraphics[width=0.75\columnwidth]{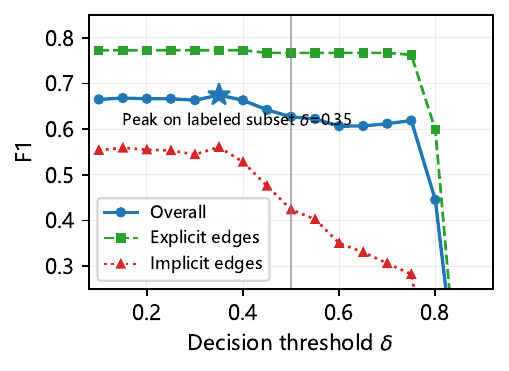}
	\caption{F1 across decision thresholds on the labeled real data subset.}
	\label{fig:threshold_sweep}
\end{figure}

The baselines include an all causal classifier and a heuristic rule. The former classifies every candidate edge as causal and provides a naive reference on the balanced annotation set. The latter requires semantic similarity above a threshold and quotation markers in comments. Results are reported in Table~\ref{tab:rq1real}. Because positive and negative labels are balanced, the all causal classifier achieves an F1 of 0.667 through 100\% recall despite having no discriminative ability. At the predefined threshold $\delta=0.5$, the proposed module achieves an F1 of 0.627, exceeding 0.498 for the heuristic rule. The threshold sweep in Fig.~\ref{fig:threshold_sweep} shows that F1 reaches 0.675 at $\delta=0.35$, indicating that the main gap arises from threshold calibration. At this threshold, explicit and latent edge F1 values are 0.773 and 0.561, respectively. The module therefore validates explicit edges more effectively, while latent edge discovery remains limited. Because $\delta=0.35$ is selected on the same labeled subset, this result is used only for threshold sensitivity analysis.

\begin{figure*}[!t]
	\centering
	\includegraphics[width=0.9\textwidth]{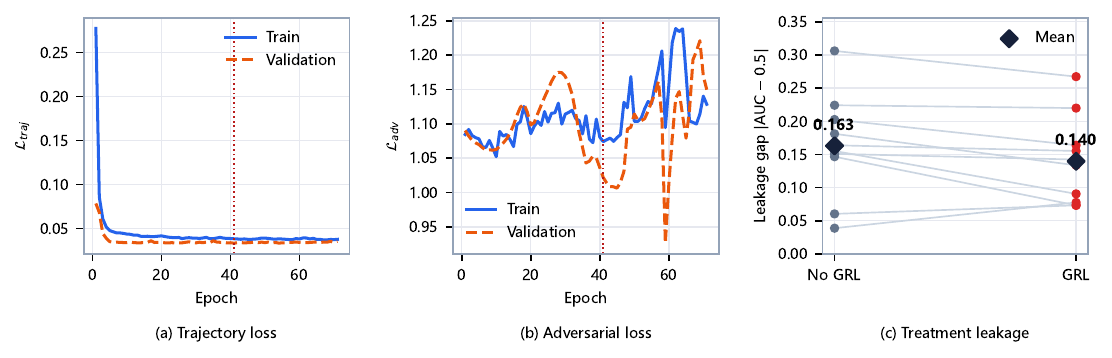}
	\caption{Training and representation balance diagnostics. (c) shows treatment information leakage from the state representation across ten repeated stratified splits. Lower values indicate that treatment status is less identifiable from $z$.}
	\label{fig:m2_training}
\end{figure*}
\subsection{Counterfactual Generation Module}\label{Sec5m2}

Table~\ref{tab:module2} reports conditional one step prediction errors over the next 24 hours under teacher forcing. The overall MSE is 0.0362, with 0.0258 for the treatment group and 0.0390 for the control group. Polarization values are concentrated and therefore yield lower errors, whereas sentiment produces the largest error. Because each step receives the true preceding trajectory, these errors are lower than those from free rollout.

\begin{table}[!t]
	\centering
	\caption{Trajectory prediction MSE under teacher forcing}
	\label{tab:module2}
	\setlength{\tabcolsep}{3.5pt}
	\begin{tabular}{lccccc}
		\toprule
		Group & Samples & Overall & Attention & Sentiment & Polarization \\
		\midrule
		Full test set & 37 & 0.0362 & 0.0251 & 0.0764 & 0.0071 \\
		Treatment & 8 & 0.0258 & 0.0215 & 0.0527 & 0.0031 \\
		Control & 29 & 0.0390 & 0.0261 & 0.0829 & 0.0082 \\
		\bottomrule
	\end{tabular}
\end{table}

\begin{table}[!t]
	\centering
	\caption{Ablation of the counterfactual generation module. Treatment permutation randomly permutes intervention encodings across samples.}
	\label{tab:ablation}
	\begin{tabular}{lccc}
		\toprule
		Variant & Test set & Treatment & Control \\
		\midrule
		Full model & \textbf{0.0499} & \textbf{0.0297} & 0.0555 \\
		w/o graph summary & 0.0529 & 0.0535 & \textbf{0.0527} \\
		w/o treatment & 0.0513 & 0.0302 & 0.0571 \\
		Treatment permutation & 0.0508 & 0.0313 & 0.0562 \\
		w/o graph \& treatment & 0.0507 & 0.0405 & 0.0535 \\
		\bottomrule
	\end{tabular}
\end{table}
\begin{table*}[!t]
	\centering
	\caption{Mean counterfactual effects with bootstrap 95\% confidence intervals. Effects are computed as factual intervention minus no intervention.}
	\label{tab:rq2}
	\small
	\begin{tabular}{lcccc}
		\toprule
		Factual intervention & Samples & Cumulative attention difference & Mean sentiment difference & Mean polarization difference \\
		\midrule
		Official announcement & 34 & $0.528\;[-0.213,1.315]$ & $0.023\;[0.007,0.039]$ & $0.034\;[0.018,0.051]$ \\
		Media guidance & 29 & $-0.474\;[-0.711,-0.211]$ & $0.034\;[0.019,0.050]$ & $-0.008\;[-0.014,-0.002]$ \\
		Account moderation & 8 & $-0.521\;[-0.916,-0.054]$ & $-0.017\;[-0.041,0.008]$ & $-0.014\;[-0.026,-0.003]$ \\
		Rumor rebuttal & 11 & $1.040\;[0.533,1.510]$ & $0.031\;[0.002,0.063]$ & $0.002\;[-0.006,0.010]$ \\
		\bottomrule
	\end{tabular}
\end{table*}
\begin{figure}[!t]
	\centering
	\includegraphics[width=0.8\columnwidth]{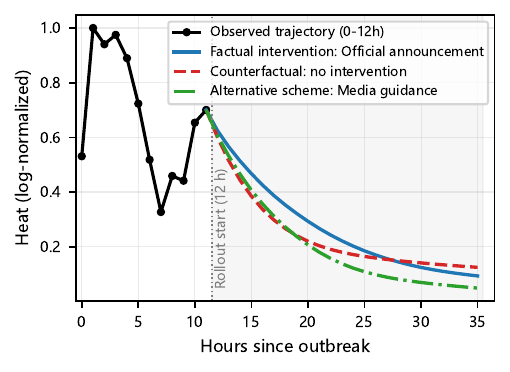}
	\caption{Normalized attention trajectories of an illustrative official announcement event under the factual intervention, no intervention, and alternative media guidance.}
	\label{fig:trajectory}
\end{figure}

Table~\ref{tab:ablation} reports free rollout MSE on the same test set, including autoregressive error accumulation. The full model achieves the lowest overall error. Removing the graph summary increases treatment group MSE from 0.0297 to 0.0535, demonstrating the contribution of propagation structure to projection for intervened events. Zeroing or randomly permuting the intervention encoding increases both overall and treatment group MSE, with the latter rising from 0.0297 to 0.0302 and 0.0313, respectively. These consistent changes show that intervention encoding provides useful incremental information. Although the treatment group contains only eight test samples, both ablations support the effectiveness of this component. Subsequent experiments fix intervention timing to control for differences in observed timing and focus on intervention type.

Fig.~\ref{fig:m2_training}(a) and (b) show the trajectory and adversarial training objectives. To test whether GRL reduces treatment assignment information in $z$, the encoder is frozen for each stratified split. A logistic regression probe is fitted using the training and validation sets, and $|\mathrm{AUC}-0.5|$ is calculated on the test set. Lower values indicate less identifiable treatment status. Across 10 repetitions in Fig.~\ref{fig:m2_training}(c), GRL reduces the mean leakage gap from $0.163\pm0.076$ to $0.140\pm0.066$, with improvements in eight runs. The bootstrap 95\% confidence interval for the paired mean change is $[-0.044,-0.003]$. Under this diagnostic, GRL reduces treatment information leakage but does not establish the removal of unobserved confounding.

\begin{table*}[!t]
	\centering
	\caption{Mean cumulative normalized attention changes after intervention replacement. Values denote alternative intervention minus factual intervention.}
	\label{tab:rq2swap}
	\begin{tabular}{lcccc}
		\toprule
		\multirow{2}{*}{Factual type} & \multicolumn{4}{c}{Replacement type} \\
		\cmidrule(lr){2-5}
		& Official announcement & Rumor rebuttal & Media guidance & Account moderation \\
		\midrule
		Official announcement & - & $+0.173$ & $-0.626$ & $-0.343$ \\
		Rumor rebuttal & $-0.243$ & - & $-0.590$ & $-0.371$ \\
		Media guidance & $+0.278$ & $+0.516$ & - & $+0.223$ \\
		Account moderation & $+0.089$ & $+0.265$ & $-0.201$ & - \\
		\bottomrule
	\end{tabular}
\end{table*}
\subsection{Question 2. Counterfactual Effects of Four Intervention Types}\label{Sec5rq2}

After evaluation on the held out test set, descriptive effects are estimated for all 82 treated events to cover every annotated intervention type. The observed history and graph summary of each event are fixed, and deterministic rollouts are generated using the factual intervention and no intervention encodings. Bootstrap sampling with replacement is performed within each intervention type. Table~\ref{tab:rq2} reports mean effects and 95\% confidence intervals. Lower attention and polarization differences are preferable, whereas higher sentiment differences indicate more positive sentiment. The intervals reflect only event sampling variability.

Media guidance shows the most consistent overall benefit by reducing attention and polarization while improving sentiment. Account moderation also benefits attention control and depolarization, whereas official announcements and rumor rebuttals mainly improve sentiment. Different actions therefore serve different objectives, and no intervention type dominates across all outcomes. Because some types contain few samples and the analysis includes training events, these results provide descriptive evidence of model conditional effects.

Fig.~\ref{fig:trajectory} presents an illustrative official announcement event with a moderate effect and clear scenario separation. Relative to no intervention, the 24 hour cumulative normalized attention differences are $+0.703$ under the factual announcement and $-0.823$ under media guidance. The same event therefore produces model estimates in opposite directions under different actions.

For all 82 treated events, the event state, intervention timing, intensity, and content encoding are fixed while only the intervention type is replaced. Mean cumulative normalized attention changes relative to the factual trajectory are then grouped by factual type, as reported in Table~\ref{tab:rq2swap}. Replacing factual media guidance with any other type increases attention. Replacing factual official announcements and rumor rebuttals with media guidance reduces attention by 0.626 and 0.590, respectively. These results broadly agree with Table~\ref{tab:rq2}, while event level heterogeneity motivates candidate comparison for each event.

\subsection{Question 3. Intervention Recommendation}\label{Sec5rq3}

Intervention recommendation is instantiated as a two objective task that reduces attention and polarization. Sentiment trajectories are still generated, but their prediction error is higher and their policy meaning depends on event context. Sentiment is therefore excluded from ranking. For event $i$, the state representation, intervention timing, intensity, and content encoding are fixed. Each of the four candidate intervention types is evaluated using $M=8$ stochastic rollouts. The mean future attention and polarization under candidate $a$ are defined as
\begin{equation}
	\bar{h}_i(a)
	=
	\frac{1}{MT_{\mathrm{future}}}
	\sum_{m=1}^{M}
	\sum_{t=1}^{T_{\mathrm{future}}}
	\hat{h}^{(m)}_{i,t}(a),
\end{equation}
\begin{equation}
	\bar{p}_i(a)
	=
	\frac{1}{MT_{\mathrm{future}}}
	\sum_{m=1}^{M}
	\sum_{t=1}^{T_{\mathrm{future}}}
	\hat{p}^{(m)}_{i,t}(a).
	\label{eq:recommendation_average}
\end{equation}
Using no intervention $a_0$ as the baseline, the candidate score is
\begin{equation}
	J_i(a)
	=
	\big[\bar{h}_i(a_0)-\bar{h}_i(a)\big]
	+
	\big[\bar{p}_i(a_0)-\bar{p}_i(a)\big].
	\label{eq:recommendation_score}
\end{equation}
Attention and polarization are normalized to $[0,1]$, allowing an equal weight sum. A positive $J_i(a)$ indicates a model estimated improvement over no intervention. Nondominated sorting first produces the Pareto frontier, from which the candidate with the highest $J_i(a)$ is selected.

\begin{figure}[!t]
	\centering
	\includegraphics[width=0.8\columnwidth]{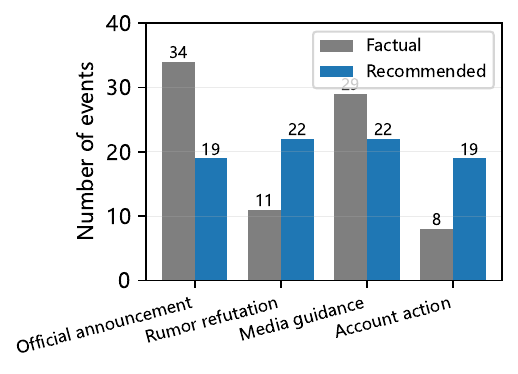}
	\caption{Factual and recommended intervention type distributions across 82 treated events.}
	\label{fig:intervention_dist}
\end{figure}

Among 82 treated events, the recommended type matches the factual type for 13 events, corresponding to 15.9\%. The model estimated gain over the factual intervention has a mean of $0.2215$ and a median of $0.1703$. As shown in Fig.~\ref{fig:intervention_dist}, media guidance and rumor rebuttal are each recommended for 22 events, while official announcements and account moderation are each recommended for 19. This distribution is more balanced than the factual distribution. The low agreement only indicates a difference between model rankings and historical choices. Because historical choices are not optimal strategy labels, agreement does not measure recommendation accuracy.

Among 296 control events without verified governance interventions, the highest scoring candidate is media guidance for 86 events, account moderation for 84, official announcements for 67, and rumor rebuttal for 59. The highest candidate score has a mean of $0.2167$ and a median of $0.2169$. At least one candidate outperforms no intervention for 236 events, corresponding to 79.7\%. No candidate outperforms no intervention for the remaining 60 events, for which no intervention should remain available. These results arise from comparisons within the same generative model and serve only as candidate screening signals rather than evidence of deployment gains.

\section{Conclusion}\label{Sec6}
A counterfactual public opinion projection framework was proposed to integrate causal structure discovery, counterfactual trajectory generation, and strategy recommendation. The causal discovery module recovered triggers between semantically dissimilar content and outperformed heuristic rules on the labeled real data subset. GRL reduced treatment information leakage from event representations. Real data estimates showed heterogeneous effects across attention, sentiment, and polarization, with media guidance providing the most consistent benefits. Intervention recommendation therefore required event specific multiobjective tradeoffs. Ablation results further supported the utility of structural summaries and intervention encodings. Existing public opinion datasets generally lack governance intervention annotations. A Weibo event dataset with such annotations is therefore constructed. However, platform access constraints and limited collection coverage restrict its scale, and the current experiments provide only initial evidence of framework feasibility. Future work focuses on collaboration with relevant institutions to obtain larger and more representative data under regulatory compliance and to further evaluate generalizability and stability.

\bibliographystyle{IEEEtran}
\bibliography{references}
\end{document}